\documentclass[doublecol]{epl2}

\usepackage[T1]{fontenc}
\usepackage{amsmath,amssymb}
\usepackage{xcolor}
\usepackage{hyperref}

\definecolor{darkblue}{rgb}{0,0,0.5}
\definecolor{orcidgreen}{RGB}{166,206,57}
\definecolor{revisioncolor}{RGB}{102,42,42}

\newcommand{\faOrcid}{\textcircled{\scriptsize iD}}
\newcommand{\orcidicon}[1]{%
  \href{https://orcid.org/#1}{%
    \textcolor{orcidgreen}{\raisebox{-0.10ex}{\faOrcid}}%
  }%
}

\hypersetup{
  pdfstartview={FitH},
  pdftitle={Finite-source magnification residuals from a cNFW-inspired perturbation of an effective spinning black hole binary lens},
  pdfauthor={Mohsen Fathi},
  pdfsubject={Finite-source magnification in an effective spinning binary lens with a cNFW-inspired radial perturbation},
  pdfkeywords={black holes, binary lenses, spin, dark matter, caustics, finite-source lensing},
  colorlinks=true,
  breaklinks=true,
  linkcolor=darkblue,
  citecolor=darkblue,
  urlcolor=darkblue
}

\title{{Finite-source magnification residuals from a cNFW-inspired perturbation of an effective spinning black hole binary lens}}
\shorttitle{{Finite-source residuals in a perturbed binary lens}}

\author{
  Mohsen Fathi
  \orcidicon{0000-0002-1602-0722}
  \thanks{E-mail: \email{mohsen.fathi@ucentral.cl}}
}
\shortauthor{M. Fathi}

\institute{
  Centro de Investigaci\'on en Ciencias del Espacio y F\'isica Te\'orica
  (CICEF), Universidad Central de Chile, La Serena 1710164, Chile
}


\abstract{{We ask whether a cored-NFW-inspired radial perturbation leaves a finite-source magnification residual after refitting the best effective spinning vacuum binary. We compute the source flux by direct image-plane integration and vary the lens scale, separation, mass ratio, two spin magnitudes, two projected spin directions, and one phase shift. After refitting, the caustics and magnification curves are almost identical and the phase shift is nearly zero. Still, a structured residual remains. At the highest resolution, the full-orbit rms residual is $1.61\times10^{-2}$ and the maximum local residual is $4.41\times10^{-2}$. Increasing the grid from $800\times522$ to $1000\times653$ changes the full-orbit rms by only $3.39\times10^{-5}$. The residual also survives nine tests with Gaussian, uniform, and tapered sources at three RMS-size factors. Thus, a finite-source residual survives strong geometric degeneracy in this effective model. We do not claim a unique dark-matter detection or a prediction for a physically scaled binary.}}

\begin{document}
\maketitle

\section{Introduction}

Caustics are the places where a lens map changes most sharply.  When a point
source crosses a fold or a cusp, new images can appear and the formal
magnification diverges.  A real source has a finite size, so this divergence
becomes a bright but finite change in the image and light curve
\cite{SchneiderEhlersFalco1992,GaudiPetters2002Fold,GaudiPetters2002Cusp}.
{Its shape depends on the source size and brightness profile.  A
finite-source calculation is therefore needed when the difference between two
lens models is small.}

Spin can change the caustic map.  In Kerr spacetime, rotation moves the
caustic away from the optical axis and opens it into an extended four-cusped
shape \cite{RauchBlandford1994,SerenoDeLuca2008,Bozza2008}.  In the
weak-deflection limit, angular momentum also shifts images and critical
curves through a small gravitomagnetic term
\cite{Sereno2003,SerenoDeLuca2006}.  A binary adds orbital motion, and its
critical curves and caustics evolve as the components move
\cite{SchneiderWeiss1986,ErdlSchneider1993,Dominik1999,Patil2017}.
{Strong-field spin caustics have also been studied directly
\cite{Bozza2003}.  In physically scaled massive-black-hole binaries, the shape
of a self-lensing flare also depends on the emitting region
\cite{DORazioDiStefano2018,DavelaarHaiman2022,DavelaarHaiman2022PRD,Wang2026}.}

Dark matter can change the same map.  The Navarro--Frenk--White (NFW) profile
is a standard model for collisionless dark-matter halos
\cite{NavarroFrenkWhite1997}.  A compact lens inside an NFW halo can move
critical curves, change image geometry, and produce caustic metamorphoses
\cite{KaramazovEtAl2021,KaramazovHeyrovsky2022}.  A cored NFW (cNFW) branch
has also been used to build a static black hole metric with modified optical
properties \cite{Senjaya2026}.  There is, however, an important caution: an
empirical halo profile does not automatically give a self-consistent
relativistic matter solution \cite{Bolokhov2026}.
{Relativistic halo models can change black hole lensing and shadows
\cite{XavierEtAl2023}, while a flexible lens fit can absorb part of a smooth
perturbation \cite{LiuEtAl2026DMSubstructure}.  Here we use the cNFW
expression only as a phenomenological radial perturbation.  We ask one narrow
question: after refitting the vacuum binary, does this perturbation leave a
finite-source magnification residual?}

\section{Effective lens map}

We use geometrised units, $G=c=1$, and dimensionless effective lens-plane
coordinates.  At each orbital phase $\Phi$, the binary is treated as an
instantaneous map from image-plane coordinates $\boldsymbol{\theta}$ to
source-plane coordinates $\boldsymbol{\beta}$:
\begin{equation}
\boldsymbol{\beta}=\boldsymbol{\theta}-\sum_{i=1}^{2}
\left(\boldsymbol{\alpha}_{M,i}+\boldsymbol{\alpha}_{J,i}\right).
\label{eq:lens-map}
\end{equation}
We set $m_1+m_2=1$ and write $q=m_2/m_1$, so that
$m_1=1/(1+q)$ and $m_2=q/(1+q)$.  {The barycentric positions are}
\begin{align}
{\boldsymbol{\theta}_1(\Phi)}&{=d\,m_2(\cos2\pi\Phi,\sin2\pi\Phi),}\\
{\boldsymbol{\theta}_2(\Phi)}&{=-d\,m_1(\cos2\pi\Phi,\sin2\pi\Phi).}
\label{eq:barycentric}
\end{align}
{We keep $q>1$ only as a label-exchange and numerical symmetry check.
For an otherwise symmetric binary it is not a separate physical branch.}

For each component, $\boldsymbol r_i=\boldsymbol\theta-\boldsymbol\theta_i$
and $u_i^2=|\boldsymbol r_i|^2+\epsilon^2$.  The mass contribution is
\begin{equation}
\boldsymbol{\alpha}_{M,i}=A m_i F_i(u_i)\frac{\boldsymbol r_i}{u_i^2},
\label{eq:mass-deflection}
\end{equation}
where $A$ sets the lens scale and $\epsilon$ smooths the centre.  The radial
factor is
\begin{equation}
F_i(u_i)=1+g_{\rm h}[\mu(R_\star u_i)-1],
\end{equation}
with
\begin{align}
\mu(R)&=1+\frac{R}{2}\left[1-\exp\left(\eta\lambda_{\rm c}\,\Xi(R)\right)\right],\\
\Xi(R)&=\frac{1}{R+\lambda_{\rm c}}-\frac{2}{R}\ln\left(1+\frac{R}{\lambda_{\rm c}}\right).
\label{eq:cnfw-mass}
\end{align}
Equation~(\ref{eq:cnfw-mass}) is adapted from Ref.~\cite{Senjaya2026}.
{For a physical axisymmetric thin lens, the radial deflection is
proportional to the projected enclosed mass $M_{2\rm D}(<\xi)/\xi$.
Equation~(\ref{eq:cnfw-mass}) does not have that meaning here.  We use it only
as a metric-inspired radial multiplier.  Each factor $F_i$ is centred on one
component, so the model describes two local effective dressings, not one
common self-consistent halo.}

For spin, we keep the leading weak-field dipole form used in
angular-momentum lensing \cite{Sereno2003,SerenoDeLuca2006}:
\begin{equation}
\boldsymbol{\alpha}_{J,i}=\kappa_J A^2m_i^2\chi_i
\left[\frac{\boldsymbol p_i}{u_i^2}-2\frac{(\boldsymbol p_i\!\cdot\!\boldsymbol r_i)\boldsymbol r_i}{u_i^4}\right].
\label{eq:spin-deflection}
\end{equation}
Here $\chi_i$ is the dimensionless spin and $\boldsymbol p_i$ gives its
projected direction.  {The coefficient $\kappa_J$ is only an effective dimensionless
normalization; it is not a calibration to a unique Kerr binary.  At the phase
of the largest residual, a flux-weighted check gives a median
$|\boldsymbol\alpha_J|/|\boldsymbol\alpha_M|=1.5\%$ and a 99th percentile of
$5.6\%$.  Also, 99\% of the weighted flux has $u_i/(A m_i)>6.6$.  The spin
term is therefore a small correction over almost all of the relevant flux.}

The cNFW benchmark uses $q=1$, $d=0.34$, $A=0.032$, $\epsilon=0.022$,
$\lambda_{\rm c}=10$, $\eta=0.0025$, $g_{\rm h}=6$, $R_\star=40$,
$\kappa_J=0.18$, $\chi_1=\chi_2=0.85$, and a projected spin angle of
$25^\circ$.  These are dimensionless parameters of an effective map.  We do
not assign a unique physical mass or distance scale.  The model includes
neither radiation reaction nor an exact binary Kerr geometry.
{The baseline finite source is Gaussian with $\sigma=0.015$.  If a
physical scaling is assigned later, the light-propagation and lensing times
must be short compared with the orbital-evolution time.  We do not test this
condition for a specific astrophysical binary here.}

{We find critical curves from the zero contour of
$\det(\partial\boldsymbol\beta/\partial\boldsymbol\theta)$ on a uniform
image-plane grid, using linear contour interpolation.  We then map them with
Eq.~(\ref{eq:lens-map}) to obtain the source-plane caustics.  After refitting,
the cNFW-inspired and vacuum curves are nearly degenerate.  Their geometry and
orbital evolution are shown in the Supplementary Material and Movie~S1.}

\section{{Finite-source magnification and vacuum fit}}

{The primary observable is the surface-brightness-integrated
finite-source magnification.  For source brightness $I_s$ centred on
$\boldsymbol\beta_s(\Phi)$, direct image-plane integration gives}
\begin{equation}
{\mu(\Phi)=
\frac{\int_{\Omega_\theta} I_s[\boldsymbol\beta(\boldsymbol\theta,\Phi)-\boldsymbol\beta_s(\Phi)]\,d^2\theta}
{\int I_s(\boldsymbol\beta)\,d^2\beta}.}
\label{eq:magnification}
\end{equation}
{The highest-resolution calculation uses a $1000\times653$ grid
over $[-0.72,0.72]\times[-0.47,0.47]$.  This finite-source magnification is
the observable used in the fit and in all quoted residuals.}

{We fit the vacuum model to the complete magnification curve.  The fit
varies $A$, $d$, $q$, $\chi_1$, $\chi_2$, $\gamma_1$, $\gamma_2$, and one
periodic phase shift.  We use broad and narrowed differential evolution,
followed by local refinements and an independent broad-seed check.  The final
parameters and search bounds are given in the Supplementary Material.  The
fitted phase shift is only $4.87\times10^{-5}$.  This small value makes the
structured residual, rather than a crossing-time shift, the relevant
diagnostic.}

\section{{Results and robustness}}

\begin{figure}[t]
\onefigure[width=0.99\columnwidth]{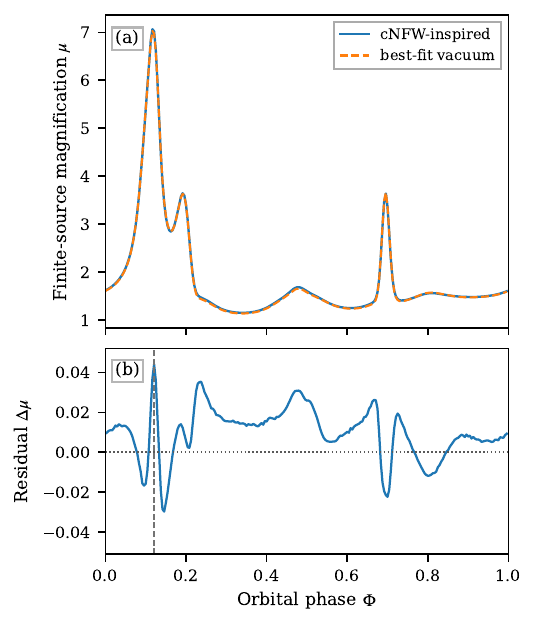}
\caption{{Finite-source magnification after refitting the vacuum
binary.  (a) The two magnification curves nearly overlap.  (b) Their residual
$\Delta\mu$.  The dashed line marks the phase of the largest local residual.
The close overlap is part of the result: the fit absorbs most of the
perturbation, but not the residual.}}
\label{fig:magnification}
\end{figure}

{Figure~\ref{fig:magnification} shows the main result.  The
cNFW-inspired and best-fit vacuum magnification curves almost overlap.  This
is the fitted degeneracy.  The lower panel shows their difference,
$\Delta\mu=\mu_{\rm cNFW}-\mu_{\rm vac}$, which is not zero and remains
structured through the orbit.  We measure it with the full-orbit rms and,
inside a fixed encounter window $\mathcal W$, the local rms and maximum
absolute residual:}
\begin{align}
{\mathcal R_{\rm orb}}&{=1.610\times10^{-2},}\\
{\mathcal R_{\mathcal W}}&{=1.560\times10^{-2},}\qquad
{\max_{\mathcal W}|\Delta\mu|=4.412\times10^{-2}.}
\label{eq:main-result}
\end{align}

{We repeated the fit at $n_x=200$, $400$, $600$, $800$, and $1000$,
with the same image-plane aspect ratio and a local vacuum refit at every
level.  From $800\times522$ to $1000\times653$, $\mathcal R_{\rm orb}$
changes by only $3.39\times10^{-5}$, or $0.21\%$ of the surviving signal.
The local rms changes by $1.70\times10^{-4}$ and the maximum residual by
$2.93\times10^{-3}$.  The point-to-caustic distance is more sensitive to the
mesh, so we do not use it as a precision observable.}

\begin{figure}[t]
\onefigure[width=0.99\columnwidth]{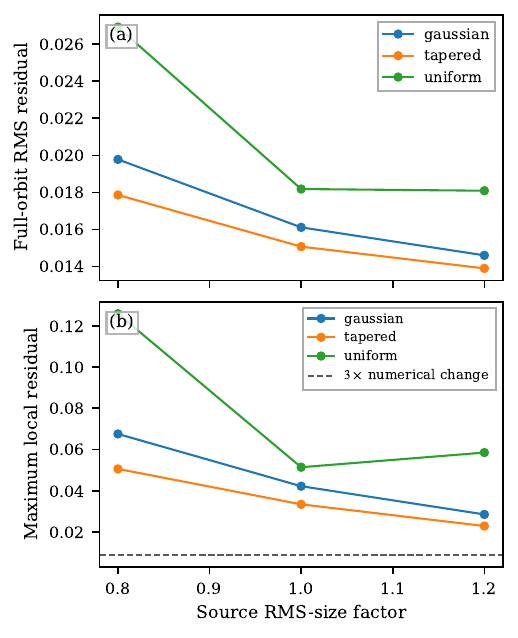}
\caption{{Source robustness after refitting the vacuum binary in
every case.  (a) Full-orbit rms residual.  (b) Maximum local residual.  The
dashed line is three times the $800\rightarrow1000$ numerical change.  The
profiles are compared at the same RMS radius.}}
\label{fig:source}
\end{figure}

{We also refit the vacuum model for Gaussian, uniform, and linearly
tapered sources at matched RMS radii of $0.8$, $1.0$, and $1.2$ times the
baseline value.  In these nine tests, the full-orbit rms ranges from $0.01390$
to $0.02692$, while the maximum local residual ranges from $0.02286$ to
$0.12581$.  The smallest values are still $410$ and $7.8$ times the
corresponding last-step numerical changes.  Figure~\ref{fig:source}
summarizes this test.}

\section{{Discussion and conclusion}}

{The main lesson is simple.  With direct finite-source magnification
and a continuous vacuum refit, the fitted phase shift is almost zero and the
caustics nearly coincide.  The quantity that survives is the structured
magnification residual.  It is stable under the numerical and source tests
reported here.}

{This residual is a model difference, not a detection significance.
A real observation would need a physical mass-distance scale, a source model,
cadence, flux calibration, photometric noise, and a treatment of intrinsic
variability.  Accretion-disc structure can also change self-lensing flare
shapes \cite{DORazioDiStefano2018,DavelaarHaiman2022,DavelaarHaiman2022PRD}.
Our dimensionless map therefore does not predict the cadence or precision
needed for an observation.}

{We also do not claim that the residual is unique to cNFW or to dark
matter.  Another smooth radial perturbation could give a similar residual,
and a flexible lens model can absorb much of a smooth perturbation after
refitting \cite{LiuEtAl2026DMSubstructure}.  Related lensed-wave studies show
the same need to separate a physical signal from model and identification
degeneracies \cite{LiuEtAl2025ML,LiuLiao2026Host}.  Our narrower result is
that the tested cNFW-inspired perturbation leaves a finite-source residual in
this effective model even after a flexible vacuum refit.  A physically scaled
binary and a self-consistent matter model are separate problems.}

\acknowledgments
The author thanks the Editor and the anonymous reviewers for their constructive comments, which significantly improved the manuscript. This work was supported by the Agencia Nacional de Investigación y Desarrollo (ANID) - Chile through FONDECYT Postdoctoral Grant No. 3260029.

\textit{Supplementary material.} The supplementary PDF gives the
source integration, fit details, numerical and source tests, geometric
near-degeneracy, approximation checks, and an illustrative image-domain
comparison.  Movie~S1 follows the critical curves, caustics, finite-source
magnification, and residual through one orbit.

\textit{Data availability.} The numerical data and code that support the findings of this study are available from the author upon reasonable request.

\bibliographystyle{eplbib}
\bibliography{caustic_dm_revised}

\end{document}